# Single-pulse time-resolved terahertz spectroscopy of sub-millisecond time dynamics

Nicolas Couture[1,2], Wei Cui[1,2], Markus Lippl[3,5], Rachel Ostic[1,2], Défi Junior Jubgang Fandio[1,2], Eeswar Kumar Yalavarthi[1,2], Aswin Vishnu Radhan[1,2], Angela Gamouras[1,4], Nicolas Joly[3,5,6], and Jean-Michel Ménard[1,2,4]

[1]Department of Physics, University of Ottawa, Ottawa, Ontario K1N 6N5, Canada
[2]Max Planck Centre for Extreme and Quantum Photonics, Ottawa, Ontario K1N 6N5, Canada
[3]Max Planck Institute for the Science of Light, Erlangen 91058, Germany
[4]National Research Council Canada, Ottawa, Ontario K1A 0R6, Canada
[5]Department of Physics, University of Erlangen-Nürenberg, Erlangen 91058, Germany
[6]Interdisciplinary Center for Nanostructured Films, Erlangen 91058, Germany

**Abstract**

Slow motion movies are not only fascinating to watch, they also allow us to see intricate details of the mechanical dynamics of complex phenomena. If the images in each frame are replaced by terahertz (THz) waves, such movies can monitor low-energy resonances and reveal fast structural or chemical transitions. Here, we combine THz spectroscopy as a non-invasive optical probe with a real-time monitoring technique to resolve non-reproducible phenomena at 50k frames per second. The concept, based on dispersive Fourier transform spectroscopy to achieve unprecedented acquisition speed of THz spectroscopy data, is demonstrated by monitoring sub-millisecond dynamics of hot carriers injected in silicon by successive resonant pulses as a saturation density is established. We anticipate that our experimental configuration will play a crucial role in revealing fast irreversible physical and chemical processes at THz frequencies with microsecond resolution to enable new applications in fundamental research as well as in industry, notably as a rapid supply chain monitoring system in high-volume manufacturing.

**Introduction**

A large class of phenomena are currently impossible to investigate since they are either too fast, non-reproducible or both. Slow motion movies and high-speed video captures help us visualize events such as the locomotion of organisms, biological processes, as well as fluid and particle dynamics in the visible and near-infrared (NIR) spectral ranges. Similar time-resolved imaging techniques in the far-infrared could provide unique insight into chemical reactions and physical processes that will not only deepen our understanding of the world around us, but also reveal insights into future technologies. Water distribution in plants[1], carrier transport in materials[2], and protein dynamics[3] could be analyzed at sub-millisecond timescales, leading to ground-breaking scientific discoveries. In industry, implementation of fast terahertz time-domain spectroscopy (THz-TDS) would provide efficient and non-invasive quality control of goods. For example, in pharmaceutical assembly lines, such a characterization technique would allow companies to determine with great accuracy the material content and thickness of the coating on their tablets, a crucial component of the solid drug delivery mechanism, without sacrificing any satisfactory tablets[4]. The current issue with THz-TDS for these applications, is that the technique mostly relies on electro-optic sampling (EOS) in electro-optic crystals[5,6]. Although EOS directly provides the full amplitude and phase information to extract the complex dielectric function of materials, it is a relatively slow point-by-point sampling process that involves mechanically scanning a NIR pulse across a THz wave. To combat this shortcoming, researchers have been attempting to decrease the scanning and data acquisition time required to retrieve the full THz transient by replacing the mechanical delay line by more sophisticated techniques such as rotary mirror arrays[7], electronically controlled optical sampling[8], asynchronous optical sampling[9], optical sampling by cavity tuning[10], and acousto-optic programmable dispersive filters[11]. These methods of THz detection have increased scanning rates from a fraction of Hz to the kHz range. Another approach to decreasing data acquisition times in THz-TDS has been to implement single-shot THz detection through frequency-, space-, and angle-to-time mapping techniques[12], spectral interferometry[13], multichannel detection[14] and an echelon mirror[15]. In fact, THz-TDS has been realized with these approaches, resolving strong magnon-magnon coupling[16] and measuring the complex dielectric function of semiconductors[17], thin metal films[18], and various THz window materials[19]. Of these methods, frequency-to-time mapping by spectral encoding is arguably the most straightforward to implement as it requires only minor modifications to a standard THz-TDS system. It is accomplished by spatially and temporally overlapping a chirped NIR detection pulse with a THz pulse in an electro-optic crystal, such as gallium phosphide (GaP) or zinc telluride (ZnTe), resulting in the time-domain waveform of the THz pulse being imprinted via nonlinear effects onto the NIR spectrum. Single-shot THz detection has been achieved with this technique by resolving THz pulses, from synchrotron[20–22] and ultrafast sources[18,23], imprinted onto a NIR pulse with a spectrometer based on CCD cameras. Nevertheless, the data collection and data transfer processes performed by the spectrometer can be drastically limited by the speed of the electronics and a large number of photodetectors, which limits the practicality of this approach. Although single-pulse THz detection has also been achieved[23–25], it has yet to be applied for THz-TDS to resolve fast dynamics in a material.

Here, we present a system capable of single-pulse THz-TDS at a repetition rate of 50 kHz based on chirped-pulse spectral encoding, dispersive Fourier transform (DFT) measurement technique, and fast detection electronics[26,27]. To explore the capabilities of our system, we monitor microsecond carrier dynamics in a silicon wafer using successive pairs of NIR pump and THz probe pulses in a transient regime. Each THz wave transmitted through the sample is time-resolved, providing phase and amplitude information, to achieve, for the first time, single-pulse THz spectroscopy of the pump-induced change in the complex dielectric function. With standard EOS technique, only the equilibrium states of the sample can be measured: the un-pumped or saturated states. With the single-pulse detection technique, we probe the sample at the repetition rate of the laser to obtain a series of measurements tracing microscopic dynamics.

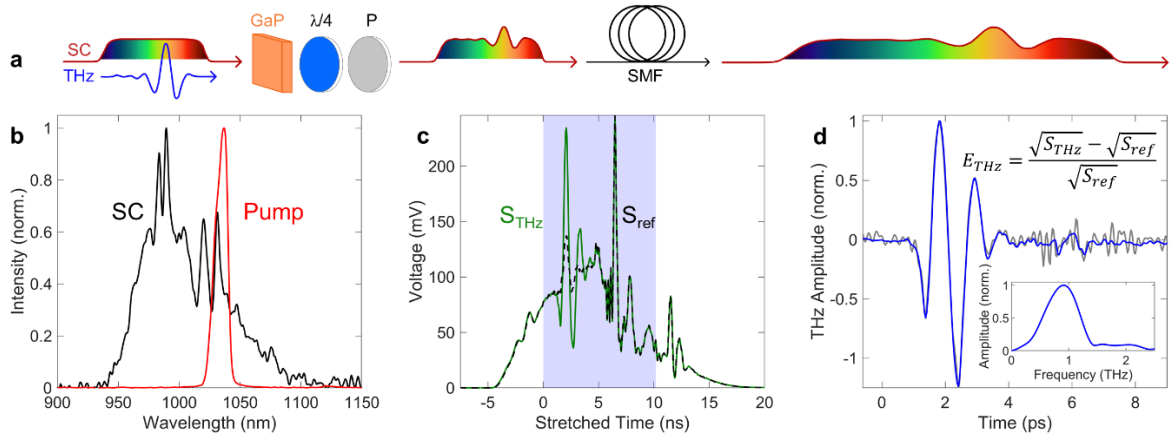

Figure 1. a) Schematic of the spectral encoding of the THz pulse onto the spectrum of the chirped supercontinuum (SC) in a 2-mm-thick GaP crystal. The modulated spectrum is passed through a quarter-wave plate ($\lambda/4$) and a polarizer (*P*) then coupled into a 2-km-long single-mode fiber (*SMF*), mapping the frequencies of the NIR pulse to the time-domain to then be resolved on an oscilloscope. b) Spectrum of the pulse injected into the short polarization maintaining fiber (*Pump*) and the output supercontinuum (*SC*), which is used for spectral encoding. c) Measured signals of the modulated ($S_{THz}$) and unmodulated ($S_{ref}$) spectra on the oscilloscope. The area highlighted in blue represents the relevant section of the spectrum which contains the THz transient. d) Extracted THz electric field ($E_{THz}$) from the curves in c) and the Fourier transform of the averaged transient (inset). The blue line is the signal averaged over 10k pulses while the grey line is a single-pulse measurement.

Using a theory based on the Drude model, we can extract the density and relaxation time of injected carriers by analyzing the complex transmission spectrum of the THz pulse. We also include in our model dynamical effects such inhomogeneous carrier distribution in the sample along the THz propagation direction, spatial diffusion, and carrier density-dependent scattering time. We believe our work lays the foundation towards resolving fast irreversible physical and chemical processes with a microsecond resolution.

**Experiment and results**

Experiments are performed with an amplified ultrafast laser source delivering 180 fs pulses (FWHM) at a central wavelength of 1030 nm, a pulse energy of 120 µJ, and a repetition rate of 50 kHz. This optical beam is split into three paths (a schematic of the experimental setup is shown in Fig. S1 of the Supplemental Material). In the first path, NIR pump pulses impinge on the sample for resonant excitation. In the second path, where most of the optical power lies, the NIR pulses are used to generate THz transients in a lithium niobate ($LiNbO_3$) crystal with a tilted-pulse-front configuration[28]. In the third path, the NIR beam is launched into a 2-meter-long polarization-maintaining fiber (PMF). Self-phase modulation and linear dispersion in the fiber yield a chirped NIR supercontinuum (SC) with ~100 nm bandwidth. This stretched pulse is then used to encode, through a nonlinear interaction process, an oscillating THz transient with an estimated temporal resolution of $\delta t = 300$ fs[29] (see Supplemental Material for details about this calculation). Figure 1a displays a schematic of the THz transient imprinted via optical rectification onto the spectrum of the chirped SC by spatially and temporally overlapping the two beams into a 2-mm-thick 110-oriented GaP detection crystal. After the nonlinear interaction process, the oscillating THz field is encoded in the polarization state of individual spectral components of the SC. To read this information with optimal sensitivity, we first use a quarter-wave plate and a linear polarizer to efficiently extract the THz-modulated signal while blocking most of the unaltered NIR beam. We then perform DFT, a single-pulse spectroscopy technique, by dispersing spectral components of the polarization-filtered SC into a 2-km-long single-mode fiber before time-resolving each nanosecond-stretched NIR pulse with a fast photodiode (Newport 1544-B) and a GHz-bandwidth oscilloscope (Tektronix MSO68B). A background signal is removed by subtracting the unmodulated SC ($S_{ref}$ in Fig. 1c) from the THz-modulated SC ($S_{THz}$ in Fig. 1c). The resulting signal is the THz transient ($E_{THz}$), which is shown in Fig. 1d along with its spectrum (inset) corresponding to the Fourier transform. The oscilloscope time axis is calibrated by comparing measurements obtained at different time intervals between the THz and detection NIR pulses, which can be accurately controlled by a delay stage.

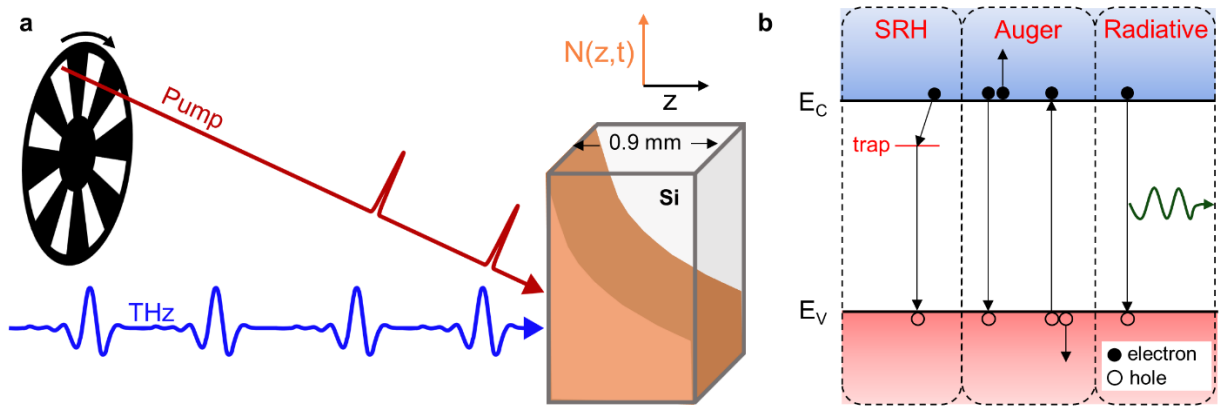

Figure 2. a) Schematic of the experimental configuration used to measure the low-energy response of optically injected carriers density $N(z,t)$ in a 0.9-mm-thick nominally undoped silicon sample. The train of NIR resonant pulses (pump) can be chopped to investigate the dynamics of pulse-to-pulse carrier injection followed by carrier relaxation. b) Schematic of the recombination mechanisms considered in simulations where red area represents the valence band ($E_V$), and blue represents the conduction band ($E_C$). Shockley-Read-Hall (SRH), Auger, and radiative recombination mechanisms are considered.

We demonstrate the capabilities of our single-pulse THz detection scheme by performing optical-pump THz-probe spectroscopy on 0.9-mm-thick nominally undoped silicon to monitor carrier dynamics. Intrinsic silicon has a relatively long carrier relaxation time of few hundreds of microseconds at low carrier densities. As a result, successive resonant pump pulses at a repetition rate of 50 kHz lead to carrier accumulation until a saturation density is reached.

We perform these experiments in two manners: i) by activating the pump and probe pulses simultaneously to observe the carrier saturation dynamics and ii) by inserting a chopper in the pump beam to monitor both the carrier accumulation and recombination dynamics, which are cyclically reproduced as the pump beam is being blocked and unblocked (Fig. 2a). We use a translation stage to adjust the time delay between the THz probe and NIR pump pulses, allowing us to adjust the THz probe any time delay before or after carrier injection by the corresponding NIR pump pulse. When the probe shortly precedes the pump pulse, spectroscopy measurements become sensitive to residual carriers with unique characteristics as they undergo about 20 µs (or the inverse of the laser repetition rate) of thermalization, diffusion and recombination. Inhomogeneous carrier distribution ($N(z,t)$) along the propagation direction of the pump ($z$) is assumed as well as Shockley-Read-Hall (SRH), Auger, and radiative carrier recombination mechanisms (Fig. 2b). We also consider spatial diffusion in our model.

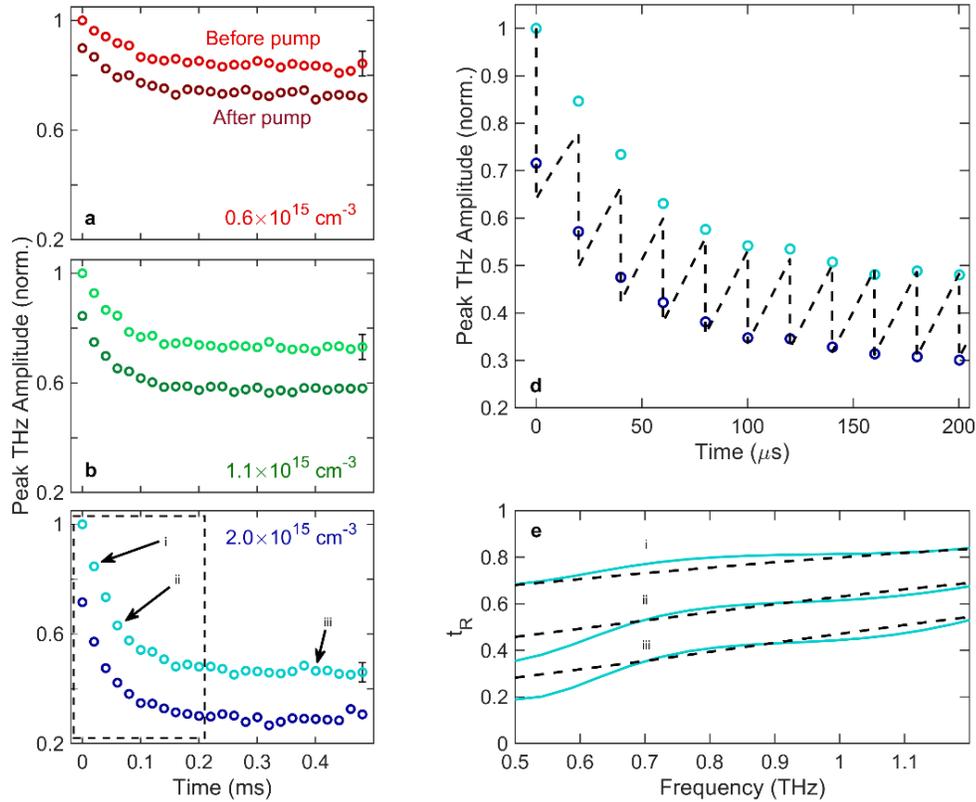

Figure 3. The peak of the THz transient as the pump pulses arrive before (dark circles) and after (light circles) the THz pulse for carrier densities of a) 0.6, b) 1, and c) $2\times10^{15}$ cm$^{-3}$. Error bars are calculated from the standard deviation over ten measurements. The area inside the dashed line box in c) is displayed in d) along with the results from theoretical calculations (dashed black line). e) Transmission of the amplitude spectra relative to the unpumped silicon ($t_R$) of the pulses labelled *i-iii* in Fig. 4c and theoretical fits. Theoretical time dynamics are calculated with the Drude model and the free carrier rate equation considering an initial recombination time of 30 µs, a density of available traps of $6 \times 10^{11}$ cm$^{-3}$, and an initial scattering time of 190 fs.

Figures 3a-c show the peak amplitude of the transmitted THz pulse, which can be used to indicate the carrier density as a function of time. We explore three pump power regimes injecting different carrier densities corresponding to $0.6 \times 10^{15}$, $1.1 \times 10^{15}$, and $2 \times 10^{15}$ cm$^{-3}$ per pulse. The experiments are repeated several times for each carrier density considered – the data presented in Fig. 3 is averaged over ten iterations. The experiment is performed as the THz probe pulse is delayed by 300 ps after the NIR pump pulse (dark coloured circles) or precedes the same pump pulse by 130 ps (light coloured circles). In our experiment, saturation carrier density is reached after 200 µs, corresponding to 10 pump pulses. At low pump pulse energy, fewer carriers are injected into the sample resulting in a lower absorption of the THz field and a higher overall transmission amplitude. Conversely, the transmission is the lowest at the highest pump pulse energy. Here we take advantage of the fact that optically injected carrier dynamics in semiconductors is a pulse-to-pulse reproducible process to average data collected over ten measurements. The error bars correspond to the standard deviation showing a relative variation of ~10% for all data points shown in Figs. 3a-c. In Fig. 3d, we plot the same data as in Fig. 3c on a shorter time scale to resolve fine details of the time-varying THz transmission during carrier accumulation from the successive NIR pump pulses. As carriers accumulate in the sample with each pump pulse, we observe an increase of the rate of carrier recombination. Our experimental results agree well with numerical calculations of photocarrier dynamics in silicon (dashed black line). In these simulations, we model the inhomogeneous injected carrier distribution by considering multiple thin slices of fixed carrier concentration across the Si sample in the direction of THz propagation ($z$). Carrier relaxation mechanisms listed in Fig. 2b are considered for each slice as well as carrier diffusion across neighbouring slices. The THz absorption of each region is modeled with the Drude model and Beer-Lambert law. Then, these results are combined to obtain the transmission amplitude through the whole sample thickness. Calculations are carried out with an initial trap-assisted effective recombination time of 30 µs, a density of available traps of $6 \times 10^{11}$ cm$^{-3}$ and an intrinsic scattering time of 190 fs. With these standard parameters[30–32], our model and experimental measurements are in good agreement. A more detailed description of the model is provided in the Supplemental Material. The real part of the transmitted THz amplitude spectrum relative to the unpumped silicon ($t_R$) of three measurements labelled *i-iii* in Fig. 3c is displayed in Fig. 3e along with the corresponding calculated transmission spectra using the model described above (dashed black lines). Here we limit our analysis to the frequency components contained within the FWHM of the THz spectrum shown in Fig. 1d. We find here also a good agreement between the model and the measurements. Note that an experimental THz scheme relying on the standard EOS detection technique could be used to characterize steady states,

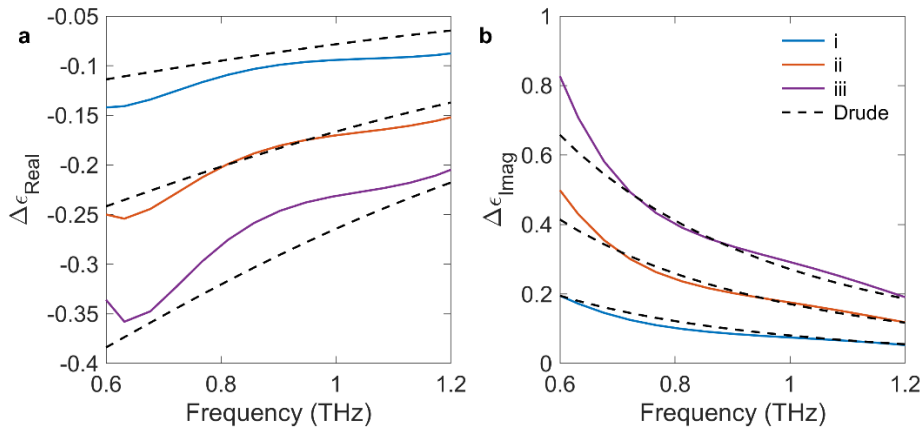

Figure 4. Extracted a) real and b) imaginary change in dielectric function calculated from the transmission spectra presented in Fig. 3e (pulses labelled *i*, *ii*, and *iii*) and the phase information of the THz pulses. Carrier distribution and scattering time (fixed at 155 fs) are used as fitting parameters for the Drude model.

corresponding here to the measurements *iii* in the saturation regime, but not the microsecond dynamics revealing transient states, such as the measurements corresponding to *i* and *ii*.

We use the complex transmission amplitude to calculate the change in the real and imaginary parts of the dielectric function, which are shown in Fig. 4a and 4b, respectively[33]. In our experiment, we can neglect the effect of the carrier density on the Fresnel transmission coefficient. As a result, the phase difference between the pumped and unpumped silicon is associated to the change in refractive index of the sample, while $t_R$, shown in Fig. 3e, is directly associated to the change in absorption. We consider the same three measurements *i*, *ii*, and *iii*, as those identified in Figs. 3c and e, but this time we fit the experimental results with a standard Drude model. The basic model is based on a homogenous carrier distribution, which is a valid approximation considering that diffusion has 20 µs after the last NIR excitation pulse to flatten the carrier distribution. We also set a non-carrier dependent scattering time for simplicity. Overall, we observe excellent fit (dashed black lines) with carrier density corresponding to $4 \times 10^{14}$, $8.5 \times 10^{14}$, $1.4 \times 10^{15}$ cm$^{-3}$ and a scattering time corresponding to 155 fs. The data presented in Fig. 4 shows that our single-pulse THz detection scheme can reliably perform spectroscopy and extract material parameters at a rate of 50 kHz, allowing us to observe dynamical changes in a system with a 20 µs resolution.

In a second type of experiment, we use a chopper wheel to block and unblock the pump beam. Carrier accumulation and recombination dynamics in Si are recorded cyclically at a frequency of 100 Hz while measuring the transmitted THz transient with the single-pulse detection technique. Figure 5a shows the peak THz amplitude over three chopper cycles and Fig. 5b provides a closer look over a single chopper cycle. This configuration not only allows us to measure the carrier injection dynamics as in Fig. 3, but also passively measures the carrier recombination dynamics when the pump pulse is blocked by the chopper (red highlighted area). On the left side of Fig. 5b, after the pump is unblocked, we have the carrier injection rate for each previously considered pump energies; and on the right side of Fig. 5b, after the pump is blocked, we have carrier recombination times which depend on the carrier density in the silicon. Our experiment yields the full recombination dynamics occurring over a millisecond time scale, while standard optical-pump THz-probe systems relying on a translation stage to incrementally change the time delay between the pump and probe are limited to nanoseconds scanning range due to the physical size of the stage. More importantly, the THz time-domain signal is sampled at a frequency of 50 kHz, corresponding to the repetition rate of the laser, allowing us to resolve fast and complex temporal dynamics. In our experiment, the excitation pulse is periodically chopped, allowing measurements to be repeated over several identical cycles. As shown in Fig. 5b, the signal-to-noise ratio corresponding to single-pulse (grey lines) data is increased when we average over three cycles (colored lines) because carrier dynamics in semiconductors is reproducible for identical optical excitation conditions. However, no averaging is intrinsically required to monitor low-energy dynamics and this is one of the main advantages of this technique. As such, we believe our system to be perfectly suited for the exploration of non-reproducible phenomena such as oxidation or combustion, or the study of chaotic dynamical systems notably arising in chemistry and biology. Finally, we believe improvements to our current setup will lead to an improved signal-to-noise ratio of single-pulse measurements allowing us to operate the system at MHz frequencies to resolve sub-µs dynamics.

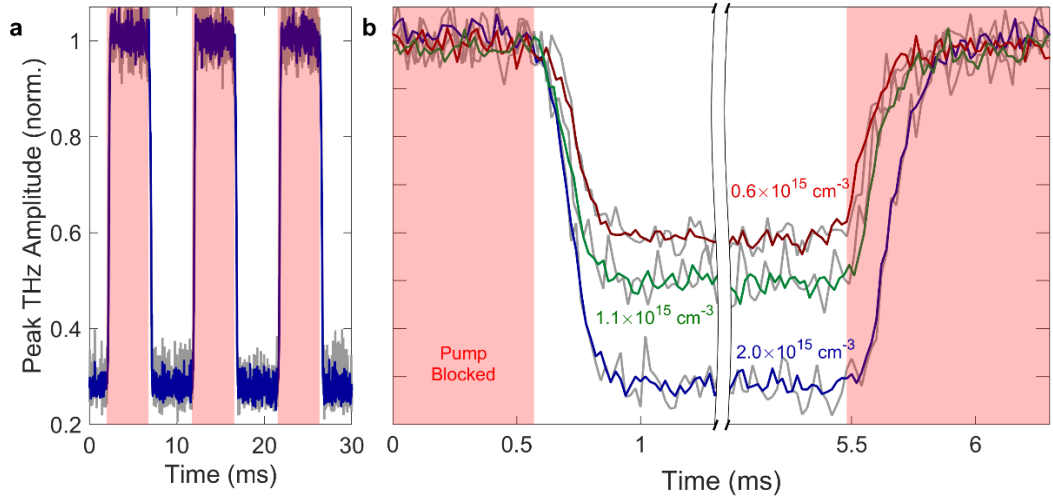

Figure 5. Optical-pump THz-probe results achieved by tightly focusing the pump beam into a mechanical optical chopper, with the pump pulse arriving before the THz probe. a) Injected carrier density is fixed at $2 \times 10^{15}$ cm$^{-3}$ and the peak THz amplitude from single-pulse THz detection is extracted over three chopper cycles. b) Peak THz amplitude extracted the same way as in a), with carrier densities of 0.6 (red), 1.1 (green), and 2 (blue) $\times 10^{15}$ cm$^{-3}$. The areas highlighted in red correspond to the situation without any pump. Coloured lines represent data averaged over 3 chopper cycles, while grey lines are single-pulse measurements.

In summary, we have presented a novel tabletop system capable of single-pulse THz spectroscopy by combining chirped-pulse spectral encoding and dispersive Fourier transform spectroscopy. With this technique, we have experimentally revealed pulse-to-pulse carrier dynamics and changes in the complex dielectric function of silicon. We have validated these experimental results with theory considering diffusion and Shockley-Read-Hall, Auger, and radiative recombination mechanisms. Although we performed our THz spectroscopy experiments at a repetition rate of 50 kHz, revealing sub-millisecond dynamics in silicon, our acquisition rate was limited only by the signal-to-noise ratio. Further development of this system will enable us to reach acquisition rates in the MHz to reveal sub-microsecond processes in systems resonant to THz frequencies. The THz-TDS technique we have presented here promises to be a powerful tool for researchers wishing to observe fast physical and chemical processes, non-reproducible phenomena, chaotic systems, and a powerful tool in industry for rapid non-invasive quality control.

**Acknowledgments**

We would like to acknowledge Testforce Systems Inc. for loaning the Tektronix MSO68B oscilloscope used in these experiments. J-MM acknowledges funding from the National Sciences and Engineering Research Council of Canada (NSERC) Discovery Grant RGPIN-2016-04797 and the Canada Foundation for Innovation (CFI) (Project Number 35269). NJ & ML acknowledge the Max-Planck Institute for the Science of Light in Erlangen for financial support.

# Supplementary material: Single-pulse time-resolved terahertz spectroscopy of sub-millisecond time dynamics

Nicolas Couture[1,2], Wei Cui[1,2], Markus Lippl[3,5], Rachel Ostic[1,2], Défi Junior Jubgang Fandio[1,2], Eeswar Kumar Yalavarthi[1,2], Aswin Vishnu Radhan[1,2], Angela Gamouras[1,4], Nicolas Joly[3,5,6], and Jean-Michel Ménard[1,2,4]

[1]Department of Physics, University of Ottawa, Ottawa, Ontario K1N 6N5, Canada
[2]Max Planck Centre for Extreme and Quantum Photonics, Ottawa, Ontario K1N 6N5, Canada
[3]Max Planck Institute for the Science of Light, Erlangen 91058, Germany
[4]National Research Council Canada, Ottawa, Ontario K1K 2E1, Canada
[5]Department of Physics, University of Erlangen-Nürnberg, Erlangen 91058, Germany
[6]Interdisciplinary Center for Nanostructured Films, Erlangen 91058, Germany

## 1. Full description and schematic of the optical-pump THz-probe experimental setup

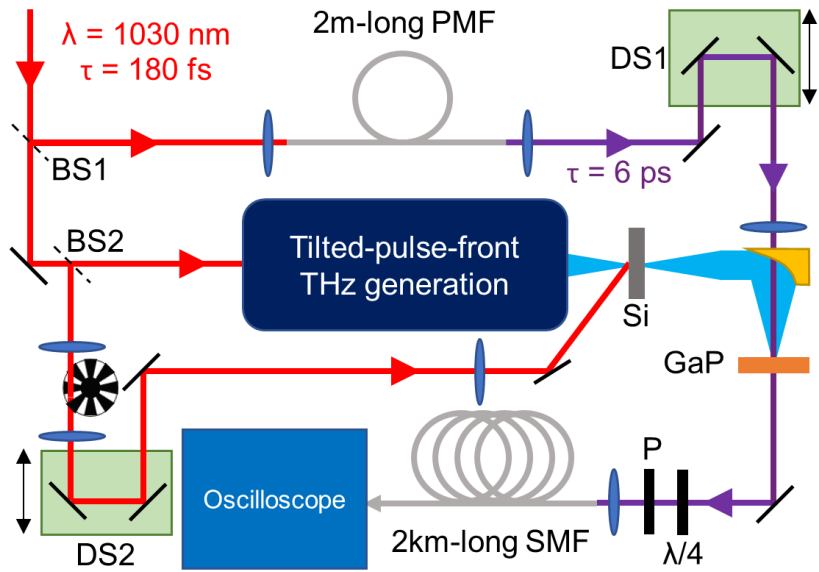

Figure S1. Schematic diagram of the optical-pump THz-probe experimental setup. BS: beamsplitter; PMF: polarization maintaining fiber; DS: delay stage; Si: silicon wafer; GaP: Gallium Phosphide; λ/4: quarter-wave plate; P: polarizer; SMF: single-mode fiber.

The output from a Yb:KGW ultrafast source (1030 nm central wavelength, 6 W average power, 50 kHz repetition rate, 180 fs FWHM time duration) (red line) is split into two initial paths with a beamsplitter (BS1, 10:90). The reflected beam, to be used as a detection pulse, is launched into a 2-meter-long polarization maintaining fiber (PMF) whose slow axis is oriented parallel to the linear polarization of the incoming pulse. Self-phase modulation and linear dispersion in the fiber yield a NIR supercontinuum (SC) with a spectral bandwidth approaching ~100 nm and a time duration of 6 ps (purple line). The temporal resolution of this method is given by $\delta t = \sqrt{T_0 T_C}$, where $T_0$ is the unchirped time duration of the SC in the Fourier transform limited case and $T_C$ is the time duration of the chirped pulse[1]. Although the spectral profile of the SC is not Gaussian (Fig. 1b of the main manuscript), a spectral bandwidth of ~100 nm (at 10 dB) is available for spectral encoding; meaning $T_0$ ~15 fs when assuming a Gaussian temporal profile. The result is a temporal resolution $\delta t = 300$ fs. The remaining power is passed through a second beamsplitter (BS2, 20:80), most of the power is used for THz generation via the tilted-pulse-front technique[2] and the remainder is used to optically pump the 0.9-mm-thick silicon wafer (Si). An optical chopper is placed in the beam pumping the silicon, placed at the focus of a 1x telescope. Delay stages adjust the temporal delay between the detection and THz pulses (DS1), and between the pump and THz pulses (DS2). The experimental setup is used in one of two manners: i), the chopper is off and the pulsed laser output is actively controlled with the pulse picker of the laser and ii), the laser remains operational with a continuous pulse train output and the chopper is turned on to activate and deactivate the pump at a frequency of 100 Hz. The THz pulse transmitted through the Si wafer and the NIR detection pulse are overlapped in a 2-mm-thick Gallium Phosphide (GaP) crystal. The time domain THz waveform modulates the spectrum of the NIR detection pulse, passes through a quarter-wave plate (λ/4) and polarizer (P), and is launched into a 2-km-long single-mode fiber (SMF). The quarter wave plate, with its fast axis aligned to the polarization of the unmodulated chirped SC, and linear polarizer, oriented such that the unmodulated chirped SC is nearly completely attenuated, filter out unmodulated NIR background light reaching the detector, thus increasing the signal to noise ratio of the measurement[1]. The SMF stretches the THz-modulated NIR pulse to the nanosecond timescale such that the spectrum of the NIR pulse is retrieved measuring the temporal intensity trace using a fast photodiode (12 GHz, 32 ps rise time) and oscilloscope (Tektronix MSO68B, 10 GHz bandwidth, 25 GSamples/s). Overall, this scheme measures the THz waveform quite accurately. The standard deviation over 10 measurements typically corresponds to 10% but can still be improved by removing external sources of noise due to air current and vibrations.

## 2. Multilayer method for evaluation of THz transmission

The transmission of the THz field through a photoexcited semiconductor is related to the dynamics and the transport of photocarriers. In bulk Si, photocarrier transport is well described by the Drude model. Moreover, the electron mobility is three times larger than the hole mobility; therefore it is assumed that charge transport is dominated by electrons[3]. Based on this assumption, the dielectric constant of Si, $\epsilon(\omega, z, t)$, can be expressed as:

$$\epsilon(\omega, z, t) = \epsilon_r + \frac{i}{\omega\epsilon_0}\sigma(\omega, z, t) = \epsilon_r + \frac{i}{\omega\epsilon_0}\frac{N(z,t)e^2\tau_s(N(z,t))}{m^*\left(1 - i\omega\tau_s(N(z,t))\right)}, \quad (1)$$

where $\omega$ is the THz angular frequency, $z$ is the distance along the THz field propagation, $t$ is time, $\epsilon_r = 11.72$ is the relative real part of the permittivity for Si at room temperature[4], $\epsilon_0 = 8.85 \times 10^{-12}$ F/m is the free-space permittivity, $\sigma(\omega, z, t)$ is the complex conductivity, $m^* = 0.2 \times 9.11 \times 10^{-31}$ kg is the electron effective mass[3], $e$ is the elementary charge, and $N(z, t)$ is the carrier density. The carrier density-dependent scattering time, $\tau_s(N(z,t))$, is expressed as:

$$\tau_s = \tau_s(N(z,t)) = \tau_0\left(1 + \sqrt{\frac{N(z,t)}{N_{ref}}}\right)^{-1} \quad (2)$$

where $\tau_0 = 190$ fs is the initial scattering time, and phenomenological parameter $N_{ref} = 10^{17}$ cm$^{-3}$ [3]. The corresponding THz power absorption coefficient can be expressed as:

$$\alpha_{THz}(\omega, t) = \frac{2\omega}{c}Im\left(\sqrt{\epsilon(\omega, z, t)}\right). \quad (3)$$

For a uniformly excited sample, the THz field transmission $T(\omega, t)$ can be expressed using the Beer-Lambert law as[5]:

$$T(\omega, t) = e^{-\alpha_{THz}(\omega,t)d/2}. \tag{4}$$

The spatial distribution of the photocarriers is not uniform upon injection as the optical pump power decreases exponentially within the sample and the photocarrier distribution is expected to change at early times after the injection due to diffusion. The multilayer method has been proven to be an accurate method to evaluate the transmission over non-uniformly distributed carriers in photoexcited samples[6]. This approach considers the photoexcited sample to consist of several layers of identical thickness where the dielectric permittivity $\epsilon(z, t)$ is constant over the considered layer. Figure S2 illustrates the partition of the photocarrier distribution into several layers. The photocarrier density in the $p^{th}$ layer is

$$N(z = pd, t = 0) = N_0 e^{-(p-1)\alpha_{pump}d} \approx N_0 e^{-\alpha_{pump}z}, \tag{5}$$

where $\alpha_{pump} = 30.2$ cm$^{-1}$ for our pump wavelength of 1030 nm[7], and $d = L/M$ is the thickness of a single layer. $L$ is the total length of the sample and $M$ is the total number of layers considered. The total THz field transmission over all layers can be expressed as the product of the THz transmission over all layers

$$T_{total}(\omega, t) = \prod_{p=1}^{M} e^{-\alpha_{THz}^{(p)}(\omega,t)d/2} \tag{6}$$

where $\alpha_{THz}^{(p)}(\omega, t)$ is the THz absorption of the $p^{th}$ layer, and $c$ is the speed of light in vacuum. The scattering time $\tau_s$ is assumed to be carrier density-dependent following Eq. 2. In this work, the value $M = 60$ was used for all simulations and corresponds to $d = 15$ µm. The derived expression of the transmission in Eq. 6 is used to evaluate the time-dependent transmission once the photocarrier dynamics over each layer are obtained.

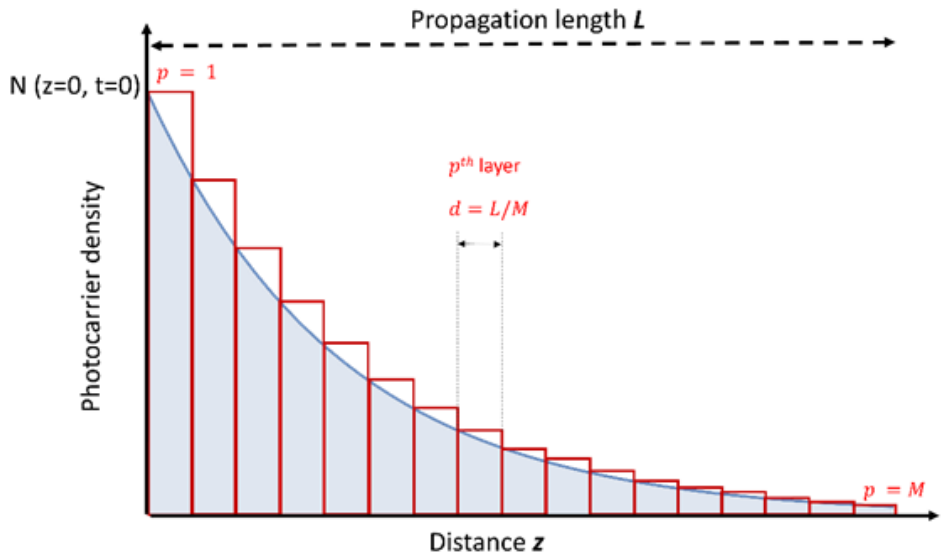

Figure S2. Spatial distribution of the photocarrier density at $t = 0$. The pump propagation distance is divided in $M$ layers of equal thickness $d$ where permittivity is assumed constant.

### 2.1. Photocarrier dynamics and spatial distribution

The equation governing free- and trapped-carrier dynamics is given by the carrier rate equation in one-dimension expressed as[8–11]:

$$\frac{dN(z,t)}{dt} = G(t) + D(N(z,t))\frac{d^2 N(z,t)}{dz^2} - \frac{N(z,t)}{\tau_{eff}(t)} \quad (7)$$

$$\frac{dN_{trap}(z,t)}{dt} = \frac{-N_{trap}(z,t) + N(z,t)}{\tau_{eff}(t)} \quad (8)$$

Where $N(z,t)$ is the density of free carriers, $G(t)$ is a photocarrier source function that generates photocarriers into the system at a rate of 50 kHz, $D(N(z,t))$ is the carrier density-dependent diffusion coefficient, $\tau_{eff}(t)$ is the trapped carrier density-dependent recombination time[11], and $N_{trap}(z,t)$ is the density of trapped photocarriers[12]. The trap-assisted recombination time is of the form:

$$\tau_{eff}(t) = \tau_R / \left(1 - \frac{N_{trap}(z,t)}{N_{trap}^{max}}\right) \quad (9)$$

where $\tau_R = 1/(\tau_{SRH}^{-1} + \tau_{rad}^{-1} + \tau_{Auger}^{-1})$ considers Shockley-Read-Hall ($\tau_{SRH}$), radiative ($\tau_{rad}$), and Auger ($\tau_{Auger}$) recombination times [11], and $N_{trap}^{max}$ is the density of available traps[12]. Equations 7 and 8 assume a uniform distribution of photocarriers in the plane of incidence of the pump beam. At early times after injection, the spatial distribution of photocarriers is expected to follow the exponential decay $N(z,t) = N(0,t)e^{-\alpha_{pump}z}$ assuming every pump photon generates an electron-hole pair. This inhomogeneous distribution is important in evaluating the total THz transmission as the penetration depth of the pump pulse $\alpha_{pump}^{-1} = 0.33$ mm is small compared to the optical propagation length of 0.9 mm. Moreover, the Si sample is periodically injected every 20 µs. By considering that photocarrier recombination is larger than the pump period, the diffusion length $L_D = \sqrt{D\tau_{decay}} = 0.26$ mm is smaller than the penetration depth. This condition entails two important consequences: the first is that of a significant impact of diffusion in the overall photocarrier dynamics, and hence on the accumulation of carriers upon successive injection, and the second implication is related to a modification of the spatial distribution of photocarriers with time as carriers diffuse in all directions. In both instances, solving Eq. 7 requires an accurate description of the spatial distribution of photocarriers at all times $t$. Previous pump-probe experiments address this issue by considering that photocarriers keep an exponential decay distribution over the probing duration[9,13]. This approximation is valid when the probing duration (a few hundred picoseconds) is very small compared to the recombination time. However, in the present case of bulk Si, the microsecond to millisecond time scale of carrier dynamics is of the same order of magnitude as the recombination time in bulk Si, and hence strengthens the need for proper modeling of $N(z, t > 0)$.

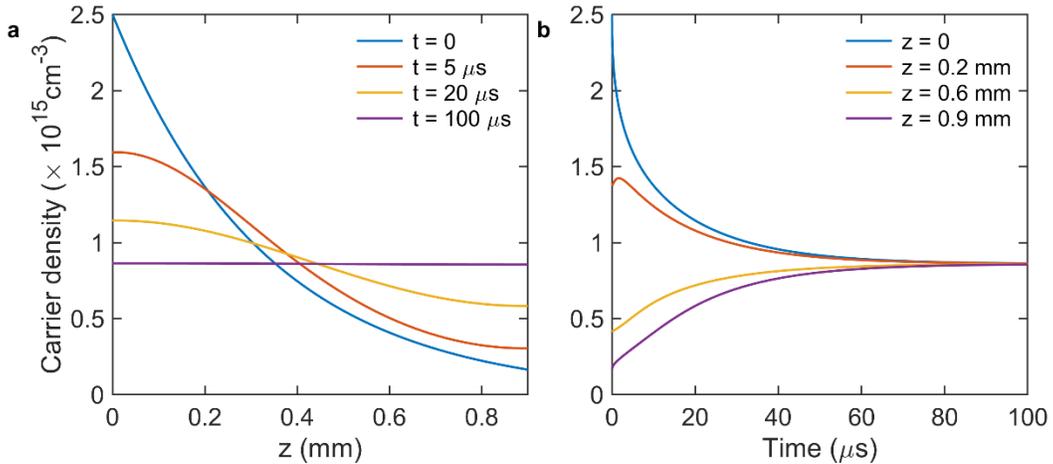

Figure S3. Time-varying carrier distribution when only considering diffusion. (a) Spatial distribution of photocarrier density at different points in time, (b) photocarrier dynamics at different points in space. Parameters considered are $N_d(z = 0, t = 0) = 2.5 \times 10^{15}$ cm³, $D = 38$ cm²/s, and $\alpha_{pump} = 30.2$ cm⁻¹.

From the exponential distribution at $t = 0$, we approximate the second derivative in the diffusion term in Eq. 7 at a given point in space ($z_i$) and time ($t_j$) with the help of central difference approximations from neighboring carrier densities ($N(z_{i+1}, t_j)$ and $N(z_{i-1}, t_j)$).

$$\frac{d^2 N(z,t)}{dz^2} = \frac{N(z_{i+1}, t_j) + N(z_{i-1}, t_j) - 2N(z_i, t_j)}{dz^2} \quad (10)$$

Figure S3a illustrates the spatial distribution of photocarriers following an injection of $2.5 \times 10^{15}$ cm$^{-3}$ carriers when calculating diffusion effects in this manner. The diffusion coefficient $D = 38$ cm$^2$/s (for initial scattering time $\tau_0 = 190$ fs at room temperature[3]) is kept fixed in Fig. S3a, but in reality is proportional to the carrier-density dependent scattering time described in Eq. 2. The calculation results found in Fig. 3 of the main manuscript and Fig. S4 utilize this carrier dependent diffusion coefficient $D(N(z,t)) = \tau_s(N(z,t))k_B T/m^*$, where $k_B$ is the Boltzmann constant and $T = 300$ K. At $t = 0$, the photocarrier distribution corresponds to an exponential decay as expected from the pump energy absorption in the sample. As time increases, this spatial distribution evolves from an exponential decay towards a uniform distribution over the entire sample thickness. Figure S3b shows the corresponding photocarrier dynamics resulting from diffusion alone at various points in space throughout the 0.9 mm thick wafer. It can be seen that carrier density at the front interface ($z = 0$) decreases rapidly shortly after excitation. On the other hand, the photocarrier density at the rear interface increases with time due to carriers diffusing from the front interface. The dynamics reach equilibrium at ~100 μs where carriers are uniformly distributed, and their density remains constant with time.

### 2.2. Photocarrier dynamics and time-varying transmission

To evaluate the carrier dynamics given by Eqs. 7 and 8 over time, we use a split-step method between carrier diffusion and recombination over time steps of 5 ns and spatial steps of 15 μm. For each point in time and space, the carrier-dependent dielectric function (Eq. 1), total THz transmittance (Eq. 6), and recombination time (Eq. 9) are calculated. The input parameters for the simulation results in Fig. 3 of the main manuscript are $\omega = 0.9$ THz (peak THz frequency in our experiments), $\tau_R = 30$ µs, $N_{trap}^{max} = 6\times10^{11}$ cm$^{-3}$. We use $N_0$, $\tau_R$, and $N_{trap}^{max}$ as fitting parameters for the experimental data. Low-doped silicon has an initial recombination time $\tau_R$ on the order of tens of microseconds[11], and the number of available traps should be much less than the injected number of carriers in a single optical pulse[12]. The values we have used for our simulations are well within these constraints.

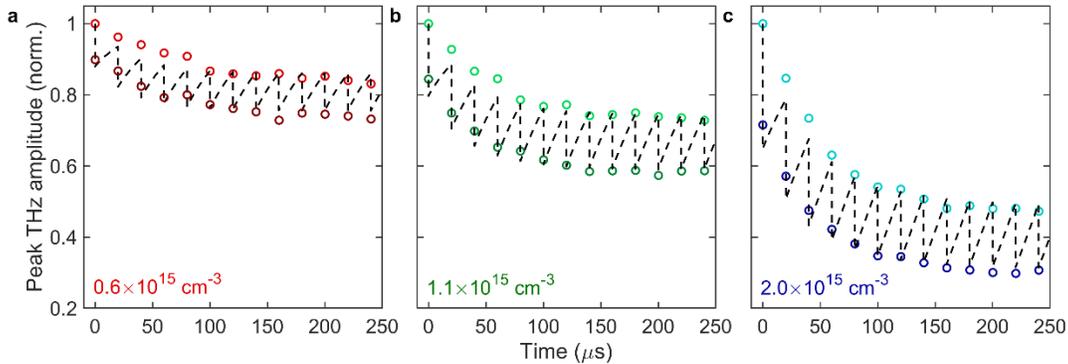

Figure S4. The peak of the THz transient as the pump pulses arrive before (dark circles) and after (light circles) the THz pulse for carrier densities of a) 0.6, b) 1.1, and c) $2 \times 10^{15}$ cm$^{-3}$. The black dashed lines represent the results from our theoretical analysis at $\omega = 0.9$ THz with the input parameters $\tau_R = 30$ μs, $N_{trap}^{max} = 6 \times 10^{11}$ cm$^{-3}$.

Here, $G(t)$ injects the carrier distribution described in Eq. 5 every 20 µs. This periodic pumping is relaxed by carrier recombination and the resulting dynamics are that of an accumulation of photocarriers in the conduction band over time. Figure S4 displays the corresponding THz transmission evaluated at the peak frequency of the THz field amplitude considering $M = 60$ layers (15 µm step size) for the three injected carrier densities considered in our experiments (Fig. 3 of the main manuscript). To calculate the theoretical THz amplitude transmission spectrum, our calculations are repeated for frequencies within the bandwidth of the THz pulse (inset Fig. 1d of the main manuscript). The numerical simulations are in good agreement with experimental results.